\input{aipcheck}

\documentclass[
    ,final            % use final for the camera ready runs
  ]
  {aipproc}

\usepackage{graphicx}
\layoutstyle{6x9}
\usepackage[figuresleft]{rotating}
\begin{document}

\title{Observe matter falling into a black hole}

\classification{
               \texttt{04.20.-q; 04.20.Cv; 04.20.Dw; 04.20.Gz; 04.20.Jb; 04.70.-s; 04.70.Bw; 95.30.Sf; 97.60.Lf}}
\keywords      {General relativity; Schwarzschild metric; Birkhoff theorem; black hole;
black hole accretion; black hole merger; frozen star; black star}

\author{Shuang Nan Zhang ({\small zhangsn@tsinghua.edu.cn})}{ address={Physics Department and Center for
    Astrophysics, Tsinghua University, Beijing 100084, China} }

\author{Yuan Liu}{ address={Physics Department and Center for
    Astrophysics, Tsinghua University, Beijing 100084, China}}

\begin{abstract}
It has been well known that in the point of view of a distant observer, all in-falling
matter to a black hole (BH) will be eventually stalled and ``frozen'' just outside the
event horizon of the BH, although an in-falling observer will see the matter falling
straight through the event horizon. Thus in this ``frozen star'' scenario, as distant
observers, we could never observe matter falling into a BH, neither could we see any
``real'' BH other than primordial ones, since all other BHs are believed to be formed
by matter falling towards singularity. Here we first obtain the exact solution for a
pressureless mass shell around a pre-existing BH. The metrics inside and interior to
the shell are all different from the Schwarzschild metric of the enclosed mass, meaning
that the well-known Birkhoff Theorem can only be applied to the exterior of a
spherically symmetric mass. The metric interior to the shell can be transformed to the
Schwarzschild metric for a slower clock which is dependent of the location and mass of
the shell; we call this Generalized Birkhoff Theorem. Another result is that there does
not exist a singularity nor event horizon in the shell. Therefore the ``frozen star''
scenario is incorrect. We also show that for all practical astrophysical settings the
in-falling time recorded by an external observer is sufficiently short that future
astrophysical instruments may be able to follow the whole process of matter falling
into BHs. The distant observer could not distinguish between a ``real'' BH and a
``frozen star'', until two such objects merge together. It has been proposed that
electromagnetic waves will be produced when two ``frozen stars'' merge together, but
not true when two ``real'' bare BHs merge together. However gravitational waves will be
produced in both cases. Thus our solution is testable by future high sensitivity
astronomical observations.
\end{abstract}

\maketitle

%%%%%%%%%%%%%%%%%%%%%%%%%%%%%%%%%%%%%%%%%%%%
%% MAINMATTER
%%%%%%%%%%%%%%%%%%%%%%%%%%%%%%%%%%%%%%%%%%%%

\vspace{-5mm}
\section{Introduction}
\vspace{-1mm}
As vividly described in many popular science writings
\cite{luminet,warp,fatal}, a distant observer ({\bf \it O}) sees a body falling towards
a BH moving slower and slower, becoming darker and darker, and is eventually frozen
near the event horizon of the BH, due to extremely strong time-dilation effect caused
by the spacetime singularity at the event horizon. Following several textbooks
\cite{mtw,Weinberg,shapiro,first_course,notes,introduction}, let's first show why the
community and general readers have acquired this view. Consider a test particle
free-falls towards a Schwarzschild BH (non-spinning and non-charged) of mass $m$ from
the last circular stable orbit of the BH, $R=\frac{6Gm}{c^2}=6m$ (here we take the
natural unit, i.e., $G=c=1$). \footnote{This starting radius is chosen because a slight
perturbation to a particle making circular Keplerian motion around the BH ar $r=6m$
will cause the particle plunge inwards, and for radius greater than $6m$ a distant
observer and an in-falling (with the test particle) observer  see almost the same
phenomena, if the particles free-falls from a larger radius.} The event horizon of the
BH is located at $r_{\rm h}=2m$. For an in-falling (with the test particle) observer
({\bf \it O'}), the in-falling velocity measured is given by
$(\frac{dr}{d\tau})^2=(1-\varepsilon^2)(\frac{R}{r}-1)$, where $\tau$ is the proper
time measured by {\bf \it O'} and $\varepsilon=-2/3$ is the energy per unit mass of the
test particle. It is clear that the test particle is seen to pass through the event
horizon with high speed. However for {\bf {\it O}} the in-falling velocity of the test
particle is given by
$(\frac{dr}{dt})^2=\frac{1}{\varepsilon^2}(1-\frac{2m}{r})^2(\frac{2m}{r}-1+\varepsilon^2)$,
where $t$ is called coordinate time, i.e., the time measured by {\bf {\it O}}. Clearly
when $r=2$, $\frac{dr}{dt}=0$, i.e., the test particle is ``frozen'' near the event
horizon of the BH, as seen by {\bf {\it O}}. For this reason an astrophysical BH formed
by in-falling matter has also been called a ``frozen'' star. The situation is depicted
in Figure 1 (left), where ``proper velocity'' and ``coordinate velocity'' are measured
by {\bf \it O'} and {\bf {\it O}} respectively.

Introducing the ``cycloid parameter'' $\eta$ by $r=\frac{R}{r}(1+\cos\eta)$, the times
taken for the test particle to reach different locations from the event horizon are
$\tau=(\frac{R^3}{8m})^{1/2}(\eta+\sin\eta)$, and
$t=2m\ln|\frac{Q+\tan(\eta/2)}{Q-\tan(\eta/2)}|+P$, where
$P=(\frac{R}{2m}-1)^{1/2}[\eta+ \frac{R}{4m}(\eta+\sin\eta)]$ and $Q=(R/2m-1)^{1/2}$,
for {\bf \it O'} and {\bf {\it O}}, respectively, as shown in Figure 1 (right). It can
be shown easily that $t$ increases exponentially as the test particle approaches to the
event horizon, as given by $r-r_{h}=P\exp(-t/2m)$, where
$\cos\eta\rightarrow\frac{4m}{R}-1$ when $r\rightarrow r_{h}$. Clearly the time
measured by {\bf \it O'} is finite when the test particle reaches the event horizon.
However for {\bf {\it O}} it takes infinite time for the test particle to reach exactly
the location of the event horizon. In other words, {\bf {\it O}} will never see the
test particle falling into the event horizon of the black hole. It is therefore
legitimate to ask the question whether in real astrophysical settings BHs can ever been
formed and grow with time, since all real astrophysical BHs can only be formed and
grown by matter collapsing into a ``singularity''.

\begin{figure}
 \begin{minipage}[t]{0.45\linewidth}
  \centering
   \includegraphics[height=0.68\textwidth]{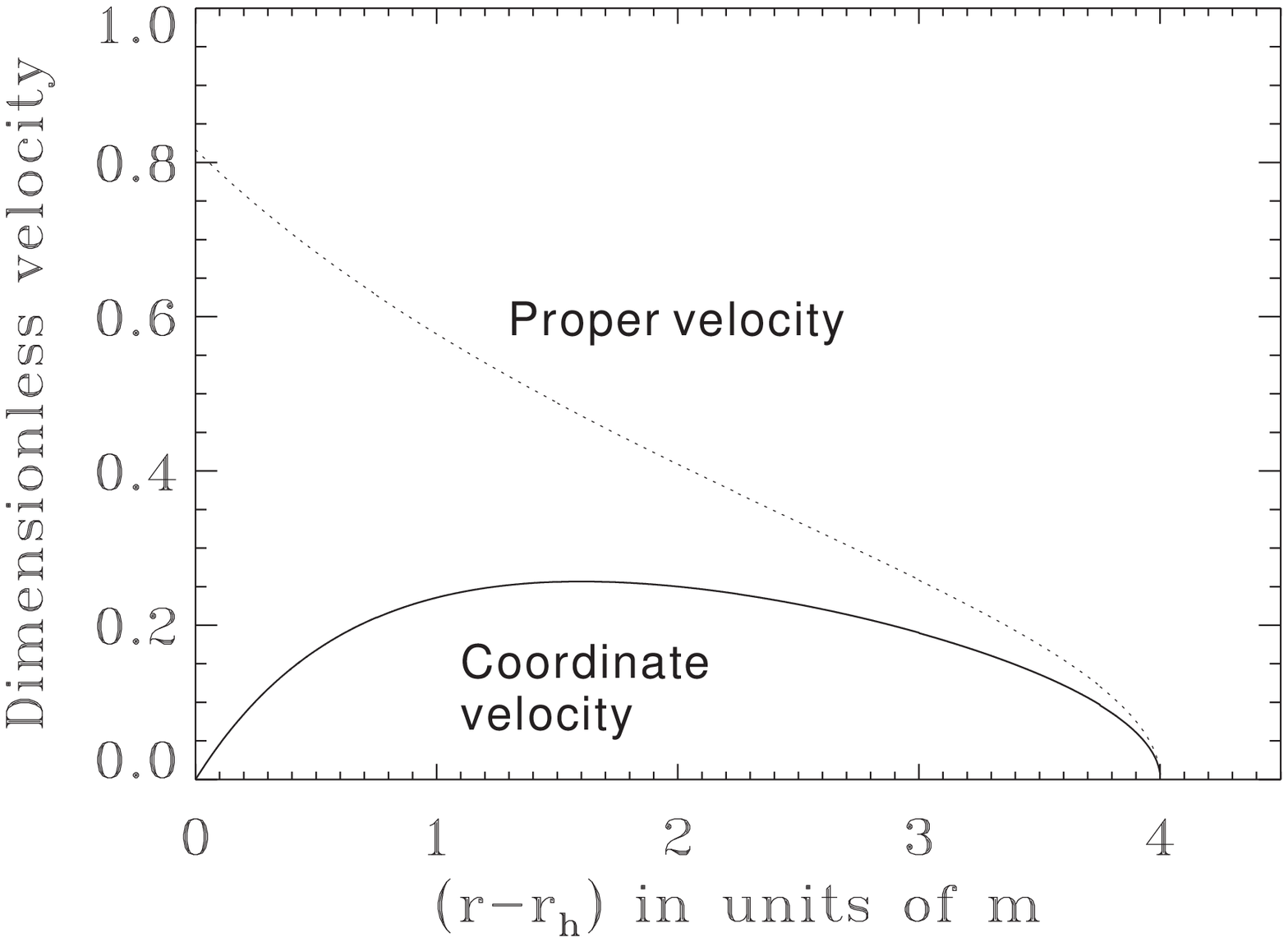}
  \end{minipage}%
  \begin{minipage}[t]{0.45\linewidth}
  \centering
   \includegraphics[height=0.7\textwidth]{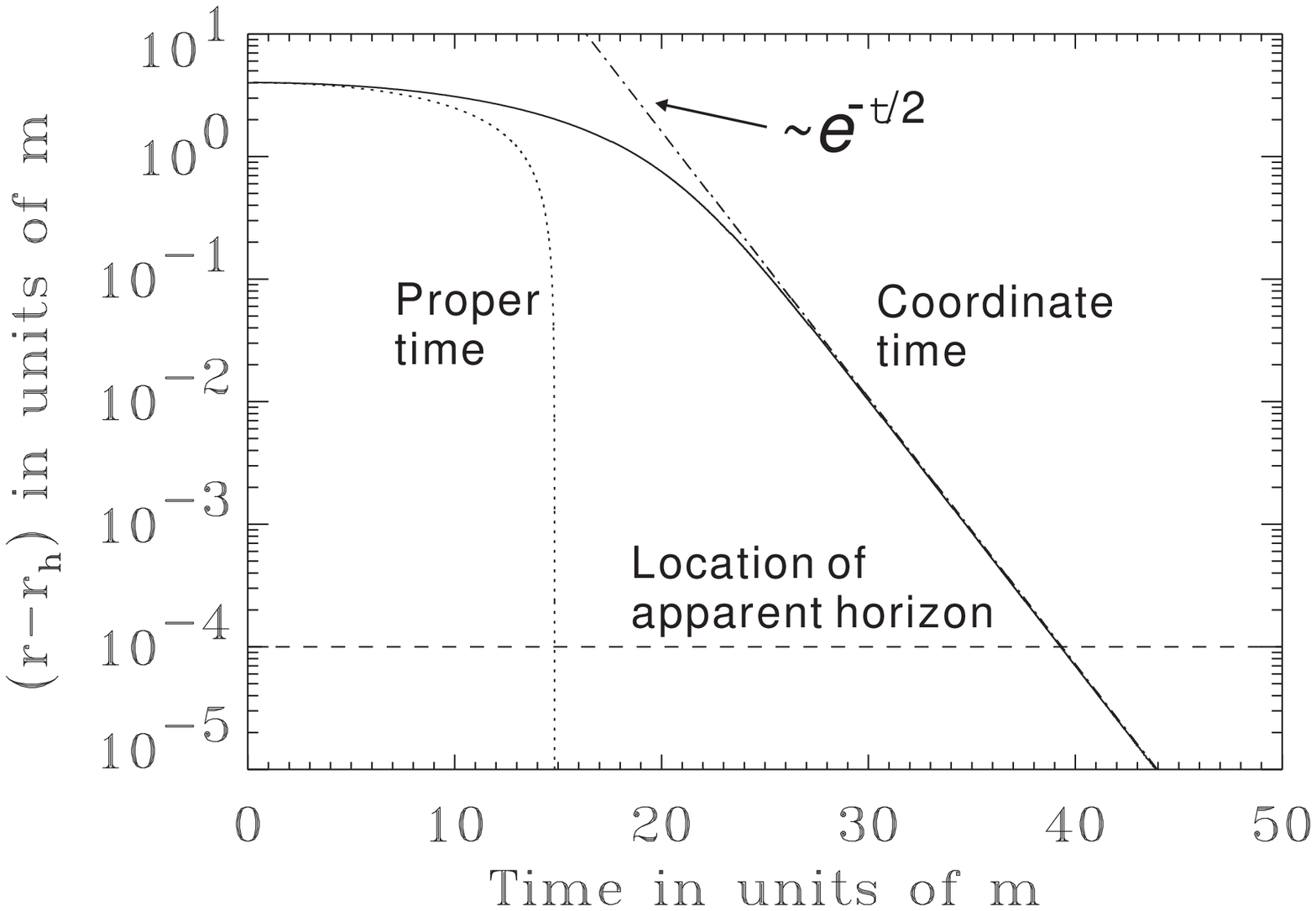}
  \end{minipage}
\vspace{-5mm}
\caption {{\bf Left:} The in-falling velocity measured by {\bf \it O'} or
{\bf {\it O}} respectively.  Here ``proper velocity'' (dotted line) and ``coordinate
velocity'' (solid line) are measured by {\bf \it O'} and {\bf {\it O}} respectively.
The test particle is seen by {\bf \it O'} to pass through the event horizon with high
speed. However, as seen {\bf {\it O}}, when $r=2m$, the ``coordinate velocity'' is
exactly equal to zero, i.e., the test particle is ``frozen'' near the event horizon of
the BH.{\hspace{1cm}} {\bf Right:} The times taken for a test particle falling towards
a BH of mass $m$, starting from $r=6m$. Here ``proper time'' (dotted line) and
``coordinate time'' (solid line) are measured by {\bf \it O'} and {\bf {\it O}}
respectively. The test particle is seen by {\bf \it O'} to pass through the event
horizon within a finite time. However the ``coordinate time'' approaches to infinity as
the test particle moves asymptotically to the event horizon, i.e., the test particle
will never move across the event horizon of the BH, as seen by {\bf {\it O}}.}
\end{figure}
Two possible answers have been proposed so far. The first one is that since {\bf \it
O'} indeed has observed the test particle falling through the event horizon, then in
reality (for {\bf \it O'}) matter indeed has fallen into the BH. However since {\bf \it
O} has no way to communicate with {\bf \it O'} once {\bf \it O'} crosses the event
horizon, {\bf {\it O}} has no way to `know' if the test particle has fallen into the
BH. The second answer is to invoke quantum effects. It has been argued that some kind
of quantum radiation of the accumulated matter just outside the event horizon
\cite{novikov}, similar to Hawking radiation, will eventually bring the matter into the
BH, as seen by {\bf {\it O}}. Unfortunately it has been realized that the time scale
involved for this to work is far beyond the Hubble time, and thus this does not answer
the question in the real world \cite{quantum,black_star}. Apparently {\bf {\it O}}
cannot be satisfied with either answer.

In desperation, {\bf {\it O}} may take the altitude of ``who cares?'' When the test
particle is sufficiently close to the event horizon, the redshift is so large that
practically no signals from the test particle can be seen by {\bf {\it O}} and
apparently the test particle has no way of turning back, therefore the ``frozen star''
does appear ``black'' and is an infinitely deep ``hole''. For practical purposes {\bf
\it O} may still call it a ``BH'', whose total mass is also increases by the in-falling
matter. Apparently this is the view taken by most people in the astrophysical community
and general public, as demonstrated in many well known text books
\cite{hawking,mtw,Weinberg,shapiro,first_course,notes,introduction} and popular science
writings \cite{luminet,warp,fatal}. However, this view is incorrect, because in real
astrophysical settings {\bf {\it O}} (in the Schwarzschild coordinates) does observe
matter falling into the BH, as we shall show as follows.

\vspace{-2mm}
\section{Solutions for a spherical shell around a BH}
\vspace{-2mm}

\begin{figure}[h]
  \begin{minipage}[t]{0.5\linewidth}
  \centering
   \includegraphics[height=0.86\textwidth]{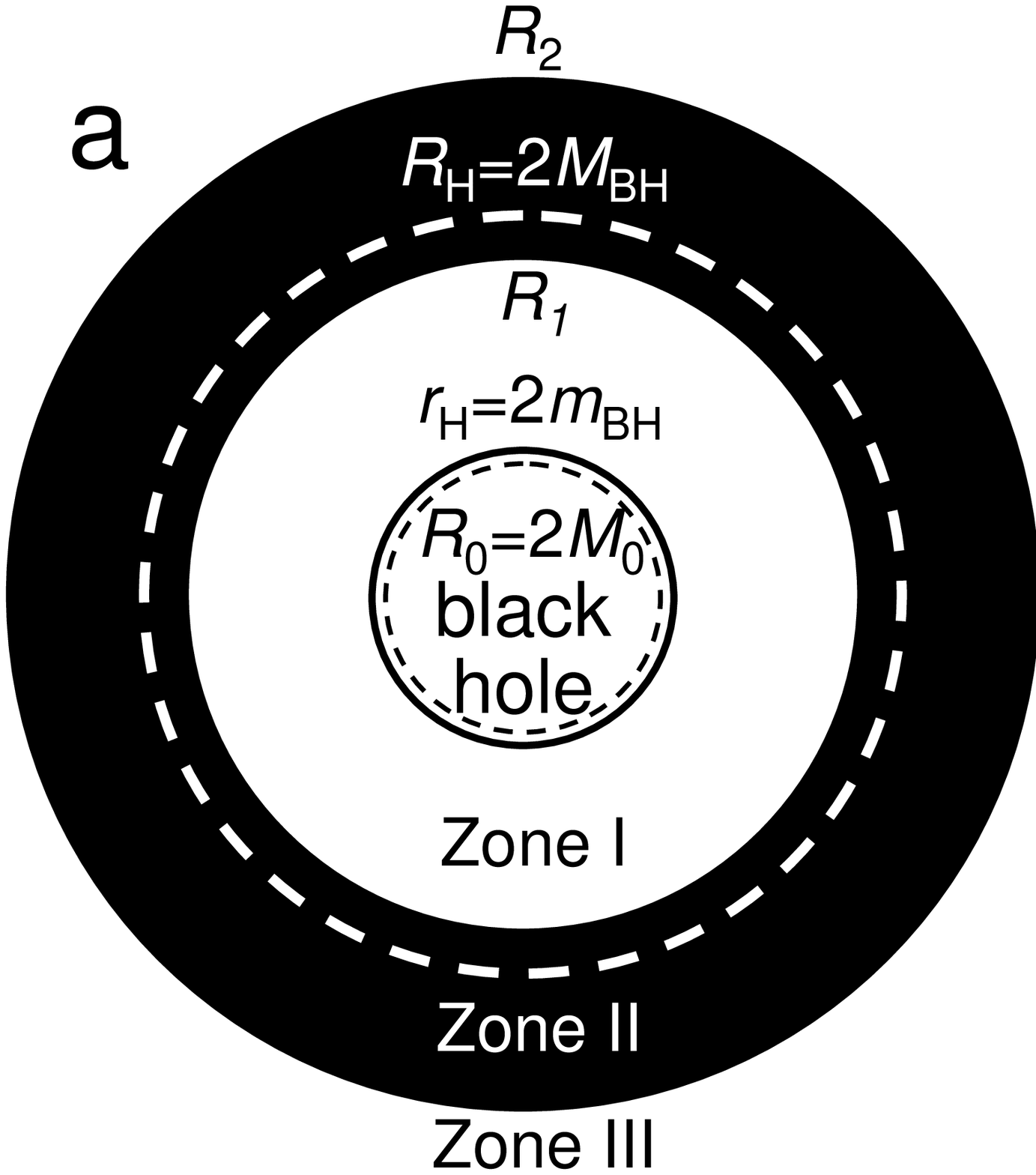}
  \end{minipage}%
  \begin{minipage}[t]{0.5\linewidth}
  \centering
   \includegraphics[height=0.86\textwidth]{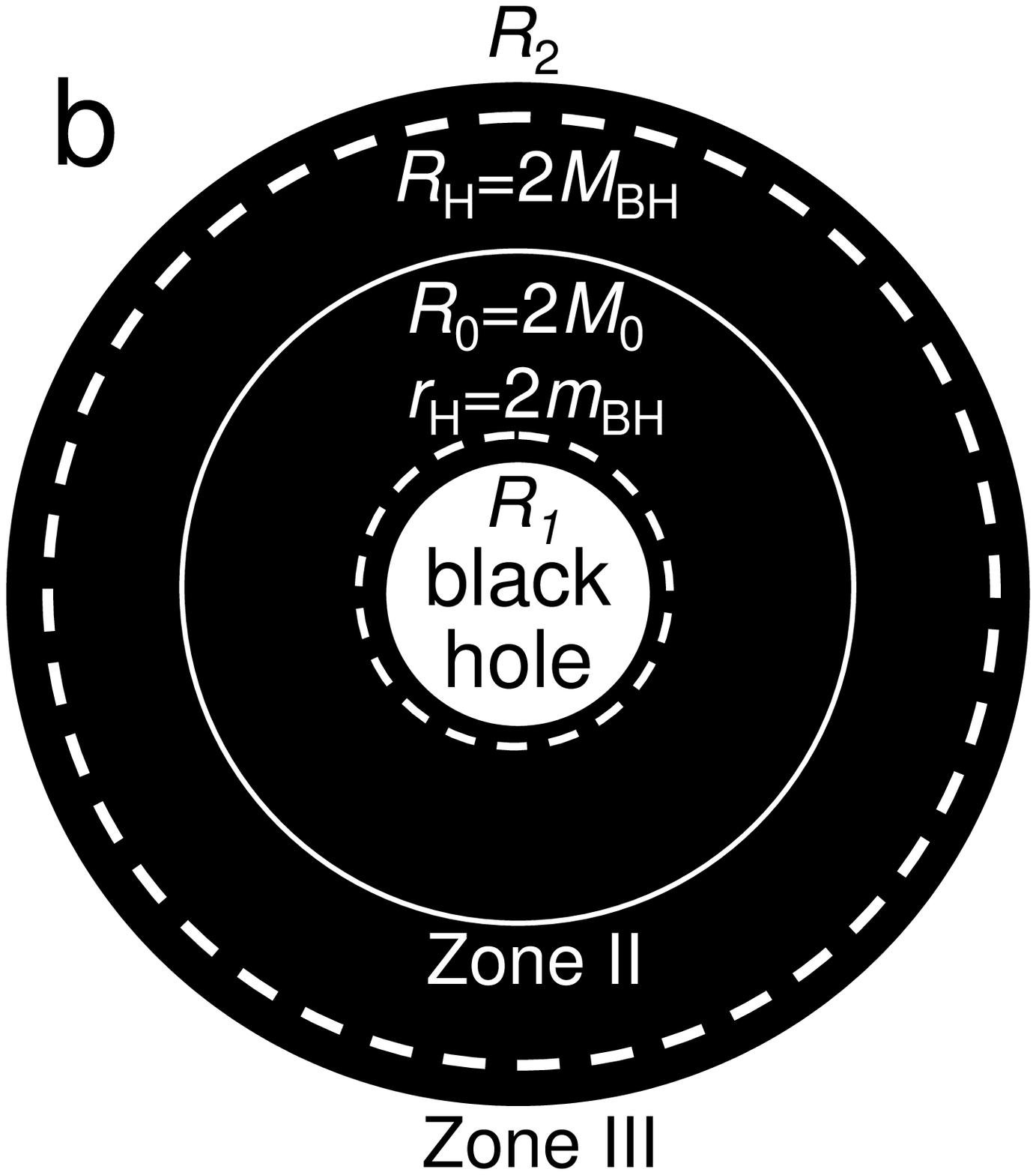}
  \end{minipage}
\vspace{-5mm}
   \caption{Illustrations of the locations of three different ``horizons'' for the system of a pressureless mass shell surrounding a
   pre-existing BH. The masses of the BH and the shell with inner and outer radii of $R_1$ and $R_2$ are $m_{\rm BH}$ and
    $m_{\rm S}$, respectively.{\hspace{1cm}} {\bf (a)} The shell is sufficiently close to, but still completely outside the BH's horizon, such that
    $R_1<R_{\rm H}=2M_{\rm BH}=2(m_{\rm S}+m_{\rm BH})<R_2$ and $R_1>r_{\rm BH}=2m_{\rm BH}$. The solution of zone {\bf III} is the Schwarzschild metric of the mass
    $M_{\rm BH}$, as given in equation (2). Since $R_{\rm H}$ is not in zone {\bf III}, there is no singularity in zone {\bf III}.
    The solution of zone {\bf II} is given in equation (3). Since $R_0=2M_0$ is not in zone {\bf II} (see text for the definition of $M_0$), there is also no singularity in zone {\bf II}.
    The solution of zone {\bf I} is given in equation (4). However, in this case, $r_{\rm BH}$ is a singularity.
    Therefore the only singularity in this system is the location of the horizon of the pre-existing BH.{\hspace{1cm}} {\bf (b)} Part of the
    shell is inside $r_{\rm BH}$. In this case, zone {\bf I} does no longer exist. The
    solutions for zone {\bf III} and zone {\bf II} take the same forms. There is still
    no singularity in zone {\bf III}, as expected. Since $R_0>R_1$, in zone
    {\bf II} there is a singularity at $R_0$, which is the apparent horizon \cite{hawking} of the new BH with a mass of $M_0$, as a consequence of the growth of the
    BH. It is important to note that in both cases the location of $r=R_{\rm H}=2M_{\rm
    BH}$ is only an apparent horizon \cite{hawking}, but neither a singularity nor the event horizon before the shell is inside $R_{\rm H}$.}
 \end{figure}
For a static \footnote{The pressureless shell cannot be maintained stationary due to
either the self-gravity of the shell or the gravity of the BH. However here we do not
study the dynamical behavior of the shell for simplicity. The solution of a shell (with
internal pressure) in dynamical equilibrium around a BH has been obtained in Ref.
\cite{shell}; the focus of that work is on how close to the BH a shell can be
maintained stationary.} and pressureless spherical mass shell around a BH as shown in
Fig.~2~(a), we assume the metric is of the form,
\begin{equation}
ds^2=B(r)dt^2-A(r)dr^2-r^2(d\theta^2+\sin^2\theta d\phi^2).
\end{equation}
For zone {\bf III}, according to the Birkhoff Theorem \cite{birkhoff,hawking,mtw} the
solution is the Schwarzschild metric for a mass $M_{\rm BH}=m_{\rm BH}+m_{\rm S}$,
where $m_{\rm BH}$ and $m_{\rm S}$ refer to the masses of the central BH and the shell
respectively. Therefore,
\begin{equation}
B_{\rm III}(r)=1-\frac{R_{\rm H}}{r},\\\\\\\\\ A_{\rm III}(r)=(1-\frac{R_{\rm
H}}{r})^{-1},
\end{equation}
where $R_{\rm H}=2M_{\rm BH}$ is the Schwarzschild radius of $2M_{\rm BH}$. Therefore
in zone {\bf III} ($r>R_2>R_{\rm BH}$), there is no singularity, even in the
Schwarzschild coordinates.

For zone {\bf II} and zone {\bf I}, the solutions can be found by applying the
continuity conditions for the metric at the inner and outer boundaries of the shell.
For zone {\bf II},
\begin{equation}
B_{\rm II}(r)=(\frac{r}{R_2-R_{\rm H}})^{\alpha-1}(\frac{1-\frac{R_0}{r}}{\alpha})^\alpha,\\\\\\\\\
A_{\rm II}(r)=\frac{\alpha}{1-\frac{R_0}{r}},
\end{equation}
where the density of matter inside the shell is assumed to have the form
$\rho=\frac{\sigma}{4\pi r^2}$ (so that the mass between any two radii inside the shell
is given by $\Delta m=\sigma\Delta r$), $\alpha=\frac{1}{1-2\sigma}$ ($\alpha>1$),
$R_0=2\alpha(m_{\rm BH}-\sigma R_1)=2M_0$, and $M_0=\alpha(m_{\rm BH}-\sigma
R_1)=\sigma(R_0-R_1)+m_{\rm BH}$. When $R_1>2m_{\rm BH}$, $R_0<2m_{\rm BH}<R_1$.
Therefore in zone {\bf II} ($R_2>r>R_1>R_0$), there is also no singularity, even in the
Schwarzschild coordinates. Note that this metric is quite different from the
Schwarzschild metric, meaning that the Birkhoff Theorem cannot be applied to zone {\bf
II}. For zone {\bf I},
\begin{equation}
B_{\rm I}(r)=(\frac{R_1-r_{\rm H}}{R_2-R_{\rm H}})^{\alpha-1}(1-\frac{r_{\rm
H}}{r}),\\\\\\\\\ A_{\rm I}(r)=(1-\frac{r_{\rm H}}{r})^{-1},
\end{equation}
where $r_{\rm H}=2m_{\rm BH}$ is a singularity. Note that this metric is also different
from the Schwarzschild metric, because the metric becomes time-dependent when the shell
starts moving. This means that the Birkhoff Theorem also cannot be applied to zone {\bf
I} directly. It can be shown easily that in the limit of a thin shell, i.e.,
$R_1\rightarrow R_2\rightarrow R$, $B_{\rm I}(r)=(\frac{R-R_{\rm H}}{R-r_{\rm
H}})(1-\frac{r_{\rm H}}{r})$, where $R$ is the radius of the thin shell. Making a
transformation $dt_{\rm I}=\sqrt{\frac{R-R_{\rm H}}{R-r_{\rm H}}}dt$ in equation (1),
the metric for zone {\bf I} becomes,
\begin{equation}
ds^2_{\rm I}=(1-\frac{r_{\rm H}}{r})dt^2_{\rm I}-(1-\frac{r_{\rm
H}}{r})^{-1}dr^2-r^2(d\theta^2+\sin^2\theta d\phi^2),
\end{equation}
which is exactly the Schwarzschild metric, but for a slower clock (since $dt_{\rm
I}>dt$) compared to the solution of the same BH without the enclosing shell; once the
shell starts moving, the clock must be re-calibrated. This can be considered as the
Generalized Birkhoff Theorem. However in most astrophysical settings, either $R\gg
R_{\rm H}$ and $R\gg r_{\rm H}$, or $R_{\rm H}\approx r_{\rm H}$; in either case,
$\frac{R-R_{\rm H}}{R-r_{\rm H}}\rightarrow 1$, thus the original Birkhoff Theorem is
still sufficiently accurate.

Therefore for the case when the shell is still completely outside the pre-existing BH,
the only singularity (when $r\ne 0$) in the Schwarzschild coordinates is at $r_{\rm
H}$, just like as if the shell did not exist. However, the metric everywhere (exterior
to, inside and interior to the shell) is influenced by the existence of the shell.

Once $R_1<r_{\rm H}$ as depicted in Fig.~2~(b), zone {\bf I} does no longer exists.
However in zone {\bf II}, $r_{\rm H}<R_0<R_{\rm H}$ and $R_0$ is the location of the
only singularity (when $r\ne 0$). In this case a temporary BH is formed with mass $M_0$
and an apparent horizon \cite{hawking} located at $R_0$. Once $R_0>R_2$, i.e., the
shell is completely within the temporary BH, the final state of a new BH is reached
with mass $M_{\rm BH}$ and an event horizon located at $R_{\rm H}$.

We note that our solutions are consistent to the graphical illustration (but without
mathematical description) of the space-time diagram for a spherical shell around a
pre-existing BH given in Ref. \cite{hawking}. Due to the page limit of the proceedings,
further details of the above solutions and their possible applications to other
astrophysical problems will be presented elsewhere \cite{liu}.
\vspace{-2mm}
\section{Shell crossing the final event horizon}
\vspace{-2mm}
 Now we turn to the question if the time measured by {\bf {\it O}} for the
shell falling into a BH is finite or infinite. First we need to clarify which BH we are
talking about, since a temporary BH will be formed during the process of a shell
falling to a pre-existing BH. Apparently the final BH with mass of $M_{\rm BH}=m_{\rm
BH}+m_{\rm S}$ should be what we are concerned with. Then the final event horizon is at
$r=R_{\rm H}=2M_{\rm BH}$. The important insight we have learnt from the exact
solutions for the shell around a pre-existing BH is that when $R_2>R_{\rm H}$, the
location of $r=R_{\rm H}$ is only the location of an apparent horizon \cite{hawking},
neither a singularity nor an event horizon. Only {\it after} $R_2$ is smaller than
$R_{\rm H}$, $r=R_{\rm H}$ will become the event horizon of the final BH and the static
clock at $r=R_{\rm H}$ will be stopped. Assuming there is some matter very far outside
the shell, i.e., the system is not placed in a complete vacuum, the static clock at
$r=R_{\rm H}$ will never be stopped {\it before} the shell is completely inside
$r=R_{\rm H}$; even the mere existence of {\bf {\it O}} can satisfy this condition.
Therefore within a finite time, {\bf {\it O}} will certainly see the shell crossing
$r=R_{\rm H}$, which would be the final event horizon of the final BH.

Next we give some numerical examples. In the case that the mass of the shell ($\Delta
m$) is much smaller than the mass of the pre-existing BH ($m$), the in-falling motion
of the shell should be very close to that of a test particle making a free-fall towards
a BH. The only difference is that the mass of the BH will be increased once the shell
reaches the horizon of the pre-existing BH. This means that a test particle will be
frozen to the horizon of the pre-existing BH, but the shell will not be frozen at the
apparent horizon \cite{hawking} at $r=R_{\rm H}$ when the shell crosses the apparent
horizon (which is not a real horizon, since in zone {\bf III} there is no singularity).
In Figure (1) (right), the location of the apparent horizon \cite{hawking} for a shell
with $(r-r_{\rm h})=10^{-4}m$ is marked. Therefore the coordinate time interval for a
test particle crossing the apparent horizon can be found easily from this plot.
Although several times longer than the proper time interval, the coordinate time
interval (called ``in-falling time'' hereafter), is still ``finite'' for all practical
physical settings as shown in the following.

Given $\frac{\Delta m}{m}=5\times10^{-3}$ (such that $(r-r_{\rm h})=10^{-4}m$), the
in-falling time is about 2 millisecond or 2.3 days for a 10 solar masses BH or a
supermassive BH of $10^{9}$ solar masses, completely negligible in the cosmic history.
Even if $\Delta m$ is decreased by a factor of $10^9$, the in-falling time is only
doubled. It is interesting to note that the above in-falling time scales are easily
accessible to current astronomy instruments; perhaps the processes of matter falling
into BHs can be indeed recorded by modern telescopes.

Let's now address briefly on the question if it is practical to see a spaceship
free-falling into a BH starting from the last stable circular orbit, such as that in
the center of the Milky Way. Assuming the mass of the spaceship is $10^3$ kg, thus
$\frac{\Delta m}{m}\approx 2\times 10^{-34}$, and consequently $t\approx 36$ minutes as
seen from the earth; the corresponding time for the clock on the spaceship is about 3
minutes. The duration of this event seems to be just adequate for the instruments on
the spaceship to conduct some experiments during the in-fall and report the results
back to the earth.

\vspace{-2mm}
\section{Conclusion and discussion}
\vspace{-2mm}

We have obtained the exact (albeit static) solutions for a spherically symmetric and
pressureless mass shell around a pre-existing BH. The metric inside or interior to the
shell is different from the Schwarzschild metric of the enclosed mass, meaning that the
well-known Birkhoff Theorem can only be applied to the exterior of a spherically
symmetric mass. Consequently the interior metric is influenced by the outside mass,
different from the Gauss Theorem for Newton's inverse-square law of gravity. This is
contrary to some (mis)understanding of the Birkhoff Theorem, e.g., in Ref. \cite{dai}.
For the region between the shell and the pre-existing BH, the metric can be transformed
to the Schwarzschild metric for a slower clock; we call this Generalized Birkhoff
Theorem. However in most astrophysical settings the original Birkhoff Theorem is still
sufficiently accurate. Another important insight from the solutions is that there does
not exist a singularity nor event horizon in the shell, i.e., the shell cannot be
frozen at the event horizon of the final BH with the total mass of the pre-existing BH
and the shell. Therefore {\bf {\it O}} can indeed observe matter falling into a BH. We
also show that for all practical astrophysical settings the in-falling time recorded by
{\bf {\it O}} is sufficiently short that future astrophysical instruments may be able
to follow the whole process of matter falling into BHs. However, more precise
calculations of the in-falling time of a shell crossing the final event horizon of the
final BH can only be done with the exact {\it dynamical} solution of a shell {\it
falling} towards a BH. Nevertheless much physical insights have been provided by our
solutions to the static shell around a BH.

Since a single ``frozen star'' appears in-different from a single ``BH'', as far as
observational consequences are concerned, one may nevertheless still ask the question:
why bother? One interesting thought experiment is to imagine what would happen if two
``frozen stars'' or two BHs merge together. It has been proposed that electromagnetic
waves will be produced when two ``frozen stars'' merge together  \cite{black_star}, if
the in-falling matter is ordinary matter rather than dark matter, but no
electromagnetic waves should be produced when two ``real'' bare BHs (no matter falling
onto the BHs) merge together. However gravitational waves will be produced in both
cases. Thus our conclusion is testable by future observations, for example, future
X-ray and gravitational wave telescopes observing the same merging events in centers of
distant galaxies.

\vspace{-2mm}
\begin{theacknowledgments}
  SNZ makes the following acknowledgments. The basic idea for this paper was sparked during a simulating
discussion in May 2007 with my former student Ms. Sumin Tang, who is now studying in
Astronomy Department, Harvard University; she also derived some formula in the first
section and wrote the code producing Figure 1. Richard Lieu of University of Alabama in
Huntsville is acknowledged for making valuable comments to the first draft of the
manuscript made in June 2007 and many follow-up discussions on this problem. I
appreciate the open-mind altitude of the organizers of this conference, mostly Drs.
Dong Lai, Yefei Yuan and Xiang-Dong Li, for allowing me to make this presentation,
which stimulated many heated debates and discussions. Many participants of this
conference, especially Drs. Kinwah Wu, Kazuo Makishima, Neil Gehrels, Masaruare Shibata
and Ramesh Narayan are thanked for interesting discussions. I thank partial funding
  supports by the Ministry of Education of China, Directional Research
  Project of the Chinese Academy of Sciences under project
  No. KJCX2-YW-T03, and by the National Natural Science Foundation of
  China (NSFC) under project No.10521001, 10733010 \& 10725313.
\end{theacknowledgments}
\vspace{-2mm}


\bibliographystyle{aipprocl} % if natbib is missing

%\bibliography{works}

\IfFileExists{\jobname.bbl}{}
 {\typeout{}
  \typeout{******************************************}
  \typeout{** Please run "bibtex \jobname" to optain}
  \typeout{** the bibliography and then re-run LaTeX}
  \typeout{** twice to fix the references!}
  \typeout{******************************************}
  \typeout{}
 }

%%%%%%%%%%%%%%%%%%%%%%%%%%%%%%%%%%%%%%%%%%%
%% The following lines show an example how to produce a bibliography
%% without the help of the BibTeX program. This could be used instead
%% of the above.
%%%%%%%%%%%%%%%%%%%%%%%%%%%%%%%%%%%%%%%%%%%

\end{document}